\documentclass[amssymb,useAMS,prd,aps,amsmath,nofootinbib,superscriptaddress,twocolumn]{revtex4}

\usepackage{color}
\usepackage{graphicx}
\usepackage{amsmath}
\usepackage[caption=false]{subfig}
\usepackage{hyperref}

\begin{document}
\title{New Method for Analysing Dark Matter Direct Detection Data}

\author{Jonathan H. Davis}
\affiliation{Institute for Particle Physics Phenomenology, Durham University, Durham, DH1 3LE, United Kingdom}

\author{Torsten En\ss lin}
\affiliation{Max-Planck-Institut fur Astrophysik Karl-Schwarzschild-Str. 1. Postfach 13 17 85741 Garching, Germany}

\author{C\'eline B\oe hm}
\affiliation{Institute for Particle Physics Phenomenology, Durham University, Durham, DH1 3LE, United Kingdom}
\affiliation{LAPTH, U. de Savoie, CNRS,  BP 110, 74941 Annecy-Le-Vieux, France}
 
 \begin{abstract}
 The experimental situation of Dark Matter Direct Detection has reached an exciting cross-roads, with potential hints of a discovery of Dark Matter (DM) from the CDMS, CoGeNT, CRESST-II and DAMA experiments in tension with null-results from xenon-based experiments such as XENON100 and LUX. Given the present controversial experimental status, it is important that the analytical method used to search for DM in Direct Detection experiments is both robust and flexible enough to deal with data for which the distinction between signal and background points is difficult, and hence where the choice between setting a limit or defining a discovery region is debatable. In this article we propose a novel  (Bayesian) analytical method, which can be applied to all Direct Detection experiments and  which extracts the maximum amount of information from the data. We apply our method to the XENON100 experiment data as a worked example, and show that firstly our exclusion limit at $90 \%$ confidence is in agreement with their own for the 225 Live Days data, but is several times stronger for the 100 Live Days data.
 Secondly we find that, due to the two points at low values of S1 and S2 in the 225 days data-set, our analysis points to either weak consistency with low-mass Dark Matter or the possible presence of an unknown background. Given the null-result from LUX, the latter scenario seems the more plausible.
 \end{abstract}
\maketitle

 \section{Introduction}

Despite convincing gravitational evidence for the existence of dark matter (DM) in our Universe (from galactic to cluster scales) its nature remains a mystery. Yet great progress has been made. In particular direct detection experiments have set progressively stronger limits on the properties of dark matter \cite{Ahmed:2011gh,Akimov:2011tj}, gaining several orders of magnitude in less than a decade for masses in the $10$ GeV to TeV range.

Several direct detection experiments have reported dark matter-like events in their data (e.g. CoGeNT \cite{Aalseth:2011wp}, CRESST-II \cite{Angloher:2011uu} and DAMA \cite{Bernabei:2010mq}), with the most recent positive result coming from the CDMS-Si experiment \cite{Agnese:2013rvf}. Such hints are in tension with the limits published by the LUX \cite{Akerib:2013tjd} and XENON100 \cite{Aprile:2012_new} collaborations. However several authors have claimed that the systematic uncertainties inherent in their analysis may provide a way of reducing such tension \cite{Davis:2012vy,Savage:2010tg,Hooper:2013cwa}. In addition if one moves beyond the most basic model of DM-quark scattering and considers e.g. inelastic scattering or isospin-violating DM, where the coupling to neutrons and protons is different, then such tension can also be greatly reduced \cite{Frandsen:2013cna,Frandsen:2011cg,Schwetz:2011xm,Chang:2010yk,Hooper:2011hd}.

Given the present situation, it is essential to exploit all the information contained in the data. In this article we propose a Bayesian approach, based on the information Hamiltonian, with a view to providing the community with a  a novel and robust interpretation of these conflicting experimental signals. This is not the first Bayesian analysis of Direct Detection data \cite{Arina:2011si}, however our method is distinct in that it extracts the maximum amount of information from the available data, by exploiting the differences between expected signal and background events. For the purpose of illustration, we will make use of data from the XENON100 calibration \cite{Aprile:2012_new,Aprile:2011hi}. This is an independent analysis of XENON100 data, and will enable us to check and also confront our new method with the collaboration's approach. {This example is also highly relevant for the LUX experiment, which works under a similar principle. }

As we will show for the case where there are signal-like points\footnote{We define ``signal-like" data as those consistent with a signal from DM, however we do not wish to make any explicit claim as to their origin, since they may also be consistent with a background interpretation.} in the data our method is particularly powerful, since one can simultaneously set an exclusion limit and define a potential signal region using Bayesian regions of credibility. This is in contrast to current analytical approaches, which usually involve methods designed only to set limits, such as the $p_{\mathrm{max}}$ method \cite{Yellin:2002xd}, or the profile Likelihood analysis with the CL$_{\mathrm{s}}$ method \cite{Aprile:2011hx}. We do not claim that our method is technically superior for all cases, however our approach is particularly transparent and easily generalised to many different data-sets.

In section \ref{sec:IH} we first introduce our method and show how to apply it to Direct Detection experimental data in general; this includes a discussion of when to set limits or claim discovery. In section \ref{sec:x100} we apply our method to data from the XENON100 experiment \cite{Aprile:2012_new,Aprile:2011hi} as a worked example and conclude in section \ref{sec:conc}.

\section{Information theory}
\label{sec:IH}

Our method is inspired from information theory, in the sense that it employs Bayesian techniques (see  \cite{Jaynes:2003} for a review) with the aim to fully exploit the different expected distributions of signal and background events. Ultimately this should either enhance the characteristics of a potential signal (and therefore the evidence for a dark matter discovery), or place stringent bounds on Dark Matter models.

{Before proceeding, we would like to clarify the distinction between this approach, and the profile Likelihood method used by e.g. the LUX \cite{Akerib:2013tjd} and XENON100 \cite{Aprile:2012_new} collaborations to set upper limits on the DM-nucleon cross section (and also by CDMS to fit to their data \cite{Billard:2013gfa}). The major difference is that our approach is Bayesian and the profile Likelihood is frequentist, and hence for example both methods have different ways of dealing with nuisance parameters. However, in most cases, with the same Likelihood function the Bayesian and profile Likelihood results should agree, and each can provide an important cross-check of the other.}

{In section \ref{sec:x100} we discuss the XENON100 experiment as a worked example, and when referring to the profile Likelihood in this context we mean specifically the Likelihood used by the XENON100 collaboration \cite{Aprile:2012_new,Aprile:2011hi} to analyse their data. Indeed, as an alternative, the LUX collaboration \cite{Akerib:2013tjd} also use a profile Likelihood method, but not necessarily the same Likelihood function as XENON100\footnote{We do not have enough information to make a statement about the Likelihood function used by the LUX collaboration.}. In the absence of any nuisance parameters, a profile Likelihood analysis performed with our Likelihood function should give similar limits to those derived in this work using a Bayesian approach. Even so, the two approaches are distinct and should be considered complimentary to each other.}

\subsection{Dividing the Data-space into a Grid}
\label{sec:method}

Our general strategy is to treat any 2D data-set effectively as an image, which we pixelate and  exploit using pattern recognition.  Said differently, we map the data contained in a 2D plot onto a 2D data-space $\Omega$. A point $x$ in this space is identified by its two coordinates $\alpha$ and $\beta$, the coordinates of the initial plot and in fact the discrimination parameters used to identify events (e.g. scintillation intensity, ionisation, phonon signals) \footnote{We have chosen a two dimensional data-space here, however our method is easily extended to data with only one parameter or several.}. 

The next key step is to then grid the data-space by pixelating it into $M$ two-dimensional bins of equal size in $\alpha  \mbox{-}  \beta$ given by $\Delta x_j = (\Delta \alpha, \Delta \beta)$ and labelled with the index $j$. If such 2D-bins are chosen to be small enough, the ability of the analysis to discriminate between signal and background will be maximised. Within a pixel $j$ at position $x_j = (\alpha_j, \beta_j)$ in the $\alpha$-$\beta$ plane there will be a certain number $n_j$ of \emph{experimental} data-points, each of which are identified by their coordinates $x^{\mathrm{data}}_i$ (with $i$ running from 0 to $N$, the total number of data-points in the whole space). For the same pixel, the \emph{theoretically expected} number of points is given by $\lambda_j = \lambda(x_j) \Delta x_j$. Hence we can compare $n_j$ to $\lambda_j$ given fluctuations in the latter, which we assume obey Poisson statistics. The function $\lambda(x_j)$ is the expected distribution of events, which constitutes the theoretical expectation of both the background and possible signal in a pixel $x_j$ \footnote{The experimental data can be thought of as a discrete sample of the theoretical distribution $\lambda(x)$.}. 

\subsection{Defining a Likelihood and Posterior}

We can now analyse the data using the method described above. The main issue is to find for which theoretical parameters is $\lambda_j$ closest to $n_j$ for all pixels $j$, within Poisson fluctuations. If there is no DM signal in the data, one expects that for the configuration where $\lambda_j$ is closest to $n_j$ that the former is equal to the theoretically expected number of background events in each pixel.

For this purpose, we will define a Poisson likelihood to describe the theoretical number of background and signal-like events in each pixel $j$. Here $\lambda_j$ represents the mean expectation value of the number of points expected in each pixel $j$. Such a Likelihood is given by,
\begin{equation}
\mathcal{P}(d|s)=\prod_{j=1}^{M}\frac{\lambda_{j}^{n_{j}}e^{-\lambda_{j}}}{n_{j}!}.
\end{equation} 
In this expression, $d$ represents the data and $s$ the signal.

To make the interpretation easier, we decompose $\lambda(x_j)$ into a DM component ${\cal{F}}(x_j)$  composed entirely of nuclear recoils (NR) and a  background component $b(x_j)$  (dominantly electronic recoils (ER), but with a possible NR component), leading to $\lambda(x_j) = {\cal{F}}(x_j) + b(x_j)$.  The predominance of the signal ${\cal{F}}(x_j)$ over the background $b(x_j)$ essentially depends on the number of signal events with respect to that from the  background, at a given location in the data-space $x_j$. Since both the number of events and the location are important, and since the location  depends on the DM mass (i.e. can be computed once for each mass), we have explicitly separated out these two contributions.
Our calculations are therefore significantly speeded up by using the decomposition:
\begin{equation}
\lambda(x_j) =  f(x_j) \ r + b(x_j) \label{lambdadecomposition}
\end{equation}
where the term $f(x_j)$ represents the signal position (or shape)  in the data-space  and $r$ its magnitude  (or intensity). For the standard picture of a non-relativistic WIMP, the interaction rate depends linearly on cross section $\sigma$, and hence $r \propto \sigma$.

The number of events is governed by the interaction cross section $\sigma$ between the Dark Matter and the nucleons of the detector. If the shape of the signal matches that of the data points (above background), then an inspection of the number of events should reveal the value of the cross section, and therefore the strength of the DM interactions.

On the other hand, if the shape does not match the data-point distribution, one can set a limit on the DM interaction cross section. In practice the finite experimental sensitivity means we can only exclude values of $\sigma$ which would lead to too large a signal. Hence it is convenient to start with a  value that is already excluded from previous experimental searches, namely  $\sigma = \sigma_0$, and decrease it until one reaches the experimental sensitivity.  For this reason we will work with the ratio $r \equiv \sigma/\sigma_{0}$  where $\sigma_0 \gg \sigma$, so that $r \equiv \sigma/\sigma_{0}$ provides us with a direct measurement of the intensity of the signal. An exclusion limit is then set by determining the smallest $r = r_{\mathrm{limit}}$ value that still leads to too many signal-like events, so that all $r > r_{\mathrm{limit}}$ are excluded, while keeping values of $r$ which the experiment is not sensitive to.

The number of expected signal events in a pixel at $x_j$ is therefore given by $f_j \ r = f(x_j) r \Delta x_j$  \footnote{We will assume here that the overall normalisation for the background is known. However in cases where this is not true one can parameterise the unknown normalisation with a nuisance parameter and associated prior, and then marginalise over it.}.
To proceed, we must now define a prior for $r$. We have no theoretical prejudice on its value and therefore consider a flat prior i.e. assign to all possible cross section values $r \in [0,1]$ the same a priori probability density function $\mathcal{P}(s(r))=\mathrm{const}$ \footnote{If we had absolutely no prejudice on the prior value of $\sigma$, we would have to take $\sigma_0 \rightarrow \infty$. However in practice we can take $\sigma_0$ to be very large but finite, such that we are confident that the probability of finding DM with this interaction strength is vanishingly small, given previous experimental knowledge.}.

We can now combine the Likelihood $\mathcal{P}(d|s)$ and prior $\mathcal{P}(s)$ into the joint data and signal probability $\mathcal{P}(d,s) = \mathcal{P}(d|s) \mathcal{P}(s)$. We will work with the information Hamiltonian,
\begin{equation}
H = - \mathrm{ln} \mathcal{P} (d,s) = \sum_{\mathrm{pixel} \, j} \left(\lambda_j - n_j \mathrm{ln} \lambda_j \right) + \dots,
\end{equation}
where the $\dots$ indicates signal-independent terms, which do not contribute to the determination of the ratio $r$. Inserting our decomposition for $\lambda(x_j)$ (cf Eq.\ref{lambdadecomposition} ) and rearranging we obtain,
\begin{equation}
H = \sum_{\mathrm{pixel} \, j} \bigg[  f_j r - n_j \mathrm{ln} \left( 1 + \frac{f_j r}{b_j}\right)   \bigg] + \dots
\end{equation}
The limit can now be taken where $\Delta x_j \rightarrow 0$, \textcolor{black}{so that each pixel} can only contain either $1$ or $0$ data-points. Hence in this limit $n_j$ tends to a delta-function and the Hamiltonian becomes
\begin{eqnarray}
H \, = \int_{\Omega} \mathrm{d}x \bigg[ f(x) r  - \mathrm{ln} \left( 1 + \frac{f(x) r}{b(x)} \right) \delta^N (x - x^{\mathrm{data}}_i) \bigg] + \dots
\label{eqn:ham_int}
\end{eqnarray}
where the $\delta$-function picks out the positions of the $N$ data-points $x^{\mathrm{data}}_i$. We define $F = \int_{\Omega} \, dx \, f(x)$, the total number of reference signal (nuclear-recoil from Dark Matter) events in the data-space calculated at $\sigma_0$.

\subsection{Setting Limits and Signal Regions}
\label{sec:sigs_lims}
With this Likelihood we are ready to look for a Dark Matter signal in our data and we now outline this process explicitly (see also \cite{Cousins:1994yw}).

As with standard $\chi^2$ methods, we seek to minimise the Hamiltonian. There  is a positive identification of a DM signal in the experimental data only when the Hamiltonian possesses a minimum. In this case the shape of the signal $f(x)$ matches the distribution of the data points, in some region of data-space where $b(x)$ is expected to be small. The strength of the DM-nucleon interaction is given by the intensity of the signal, $r_{\mathrm{best}}$,  corresponding to $\partial_r H(d,s_{\mathrm{best}}) = 0$, with $s_{\mathrm{best}}$ representing the properties of the signal that fit the data best.

To define the goodness of the fit in the standard approach, one would then consider all $r$ (or equivalently $\sigma$) values leading to $\chi^2 = \chi^2_{\mathrm{best}} + \delta $ where $\delta$ is fixed by the confidence level that one wants to have. Here we shall proceed slightly differently (but ultimately this is equivalent): we define the significance of the signal by integrating the Posterior distribution 
\begin{equation}
\centering
\mathcal{P} (s|d) = \frac{\mathcal{P}(d,s)}{\mathcal{P}(d)} \overset{\mathrm{f.p.}}{=} \mathcal{P}(d|s),
\end{equation}
over $r$, retaining in particular $r$ values around $r_{\mathrm{best}}$. 

Note that the last equality holds only for flat priors (f.p.),  and assuming that $\mathcal{P}(d) = \mathcal{P}(s)$. However, in the following we will take out the normalisation of $\mathcal{P}(d|s)$  explicitly, such that:
\begin{equation}
\centering
\mathcal{P} (s|d) =  \frac{\mathcal{P}(d|s)}{\int \mathrm{d} r \,\mathcal{P}(d|s)}.
\end{equation}

Hence in our case a discovery will be established at a  confidence level  $X$  by using the definition,
\begin{eqnarray}
\int_{r_{\mathrm{low}}}^{r_{\mathrm{best}}} \mathrm{d} r \, \mathcal{P} (s(r)|d) = \int_{r_{\mathrm{best}}}^{r_{\mathrm{up}}} \mathrm{d} r \, \mathcal{P} (s(r)|d) = X/2,
\end{eqnarray}
where the discovery region is bounded from below by $r_{\mathrm{low}}$ and from above by $r_{\mathrm{up}}$. Such a region is therefore a two-sided region of credibility, while an exclusion limit by contrast is said to be one-sided. One could also relate our two-sided Bayesian region of credibility to a frquentist confidence interval with a certain number of `sigmas', though this is only strictly possible for a Gaussian Likelihood and Posterior\footnote{In such a case the size of a the region between $r_{\mathrm{low}}$ and $r_{\mathrm{up}}$ could be directly related to the distance from the best-fit point $r_{\mathrm{best}}$ in units of the Gaussian variance i.e. a number of `sigmas'.}.

However one may find that the Hamiltonian possesses no minimum. In this case there is no value of $r$ for which the data is compatible with the signal distribution, no matter how intense this distribution becomes. One can not completely rule out Dark Matter however, since we know that our experiment has finite sensitivity, but we can set a limit, hereafter referred to as $r_{\mathrm{limit}}$, on the DM interactions.

Since the experiment is not sensitive to  DM cross section values smaller than $\sigma_{\mathrm{limit}} = r_{\mathrm{limit}} \times \sigma_0$, all $r$ values  below  $r_{\mathrm{limit}}$ are equally good (or equally bad).  Hence there is a region of the parameter space corresponding to $r< r_{\mathrm{limit}}$ where the Posterior probability $\mathcal{P} (s|d)$ is practically constant, as the experiment cannot discriminate between these values of the cross section (for a given exposure).

The allowed region below $r_{\mathrm{limit}}$ is thus characterised by $\mathcal{P} (s|d)=cst$ while the excluded region above $r_{\mathrm{limit}}$ (where one expects too much signal) is identified by a sharp cut-off in the posterior probability. To determine the exclusion limit (i.e.  $r_{\mathrm{limit}}$), we thus seek to quantify this cut-off.  We have some freedom in choosing its value: it will depend on the confidence with which we set out limit. For example to set an exclusion limit at a confidence of $Y$ (e.g. for $90 \%$ confidence we take $Y = 0.9$), we define $r_{\mathrm{limit}}$ analogously with our best-fit region, as 
\begin{equation}
\int_0^{r_{\mathrm{limit}}}  \mathrm{d} r \, \mathcal{P} (s(r)|d) = Y.
\end{equation}
By integrating the constant region of the posterior probability until the integration reaches the value that we set,  we identify $r_{\mathrm{limit}}$ and the cut-off.

Note also that for ease of calculation we tend to use the Hamiltonian in the form of,
\begin{equation}
H  = F\, r-\sum_{i=1}^{N}\ln\left(1+w_{i} r\right),
\label{eq:sigma_limit}
\end{equation}
where $i$ sums over all $N$ data-points at positions $x_i$ and $w_i$ are data weights with $w_i = f(x_i)/b(x_i)$. For setting a limit the first term in eqn. (\ref{eq:sigma_limit})  $F r$ is data-independent and gives the absolute limit in the case where no signal-like events are observed in the data, while the second term accounts for potential signal-like events present in the data, and weakens the limit.

The statistical treatment is largely similar for setting limits or claiming discovery, and our method provides a natural transition between the two, though the approach to how one thinks about regions of credibility is different in either case. Indeed both a signal region and an exclusion limit are equally valid regions of credibility, and so one may wish to highlight both if there is a hint of signal present in the data, but one wishes to remain conservative as to its interpretation.

\section{Worked Example: XENON100 Data}
\label{sec:x100}
 \begin{figure*}[t]
\centering
\subfloat{\label{fig:dists_20} \includegraphics[width=0.40\textwidth,trim=0 25 0 0,clip=True]{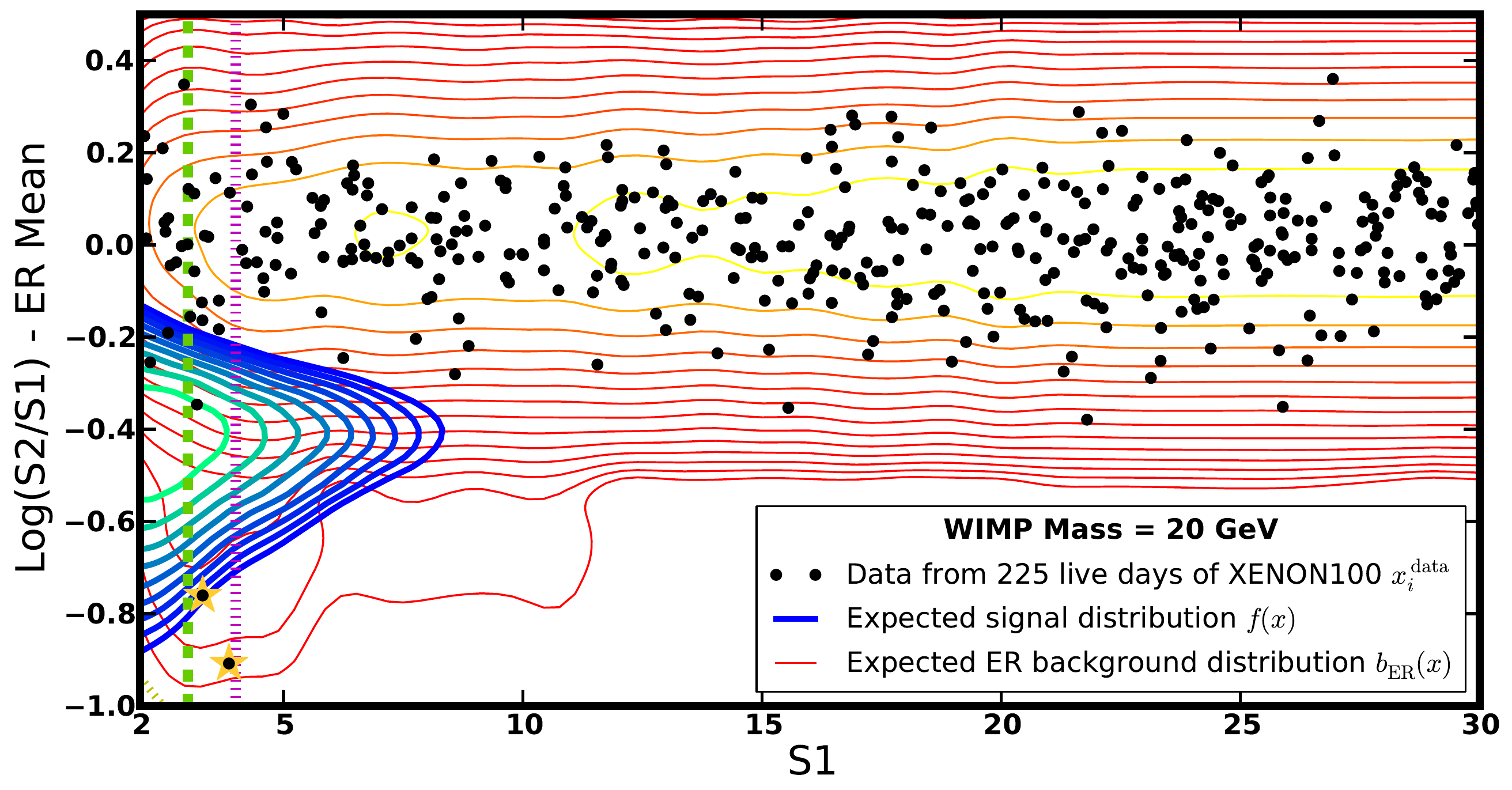}} 
\subfloat{\label{fig:dists_20_b} \includegraphics[width=0.43\textwidth,trim=0 25 0 0,clip=True]{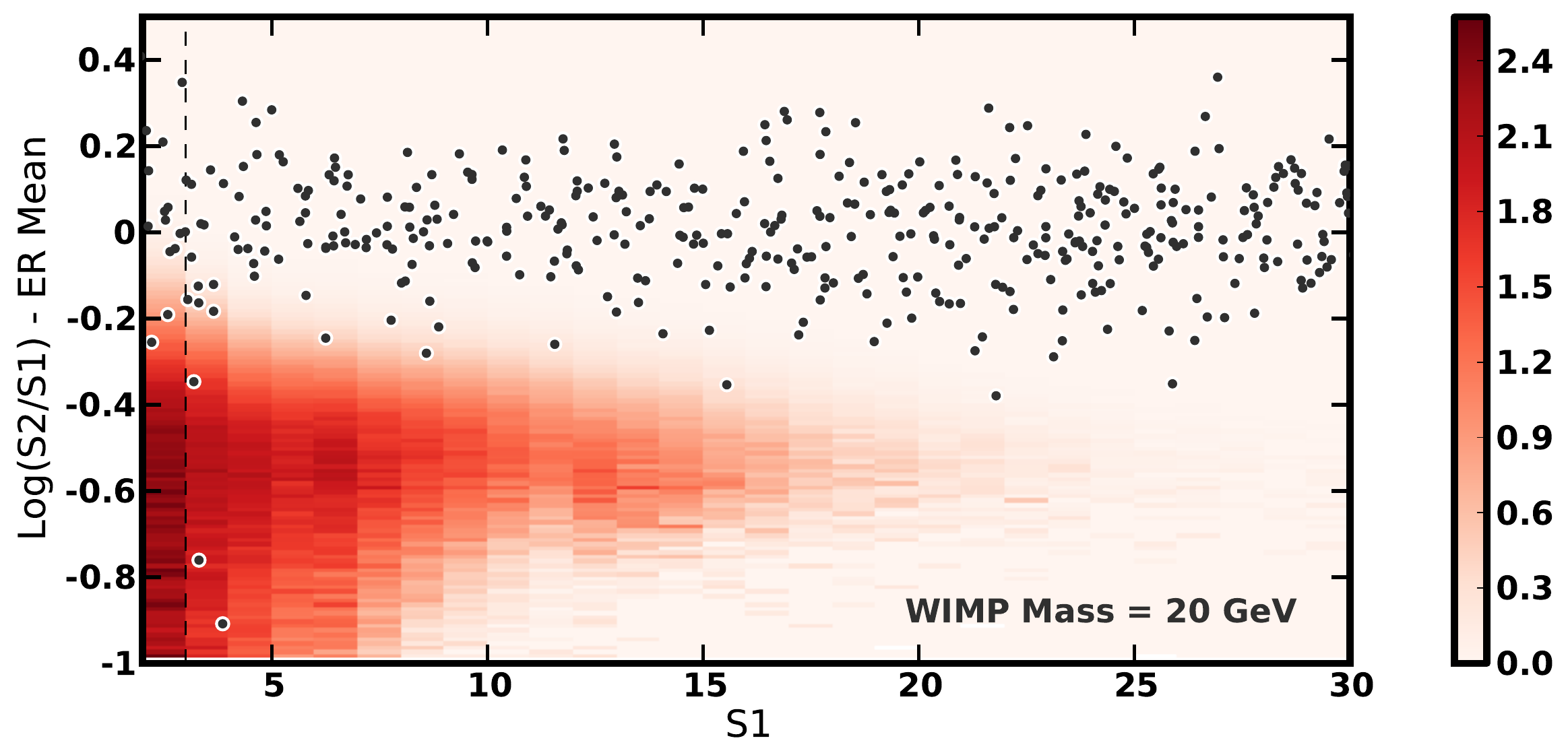}} \\
\subfloat{\label{fig:dists_100} \includegraphics[width=0.40\textwidth,trim=0 25 0 0,clip=True]{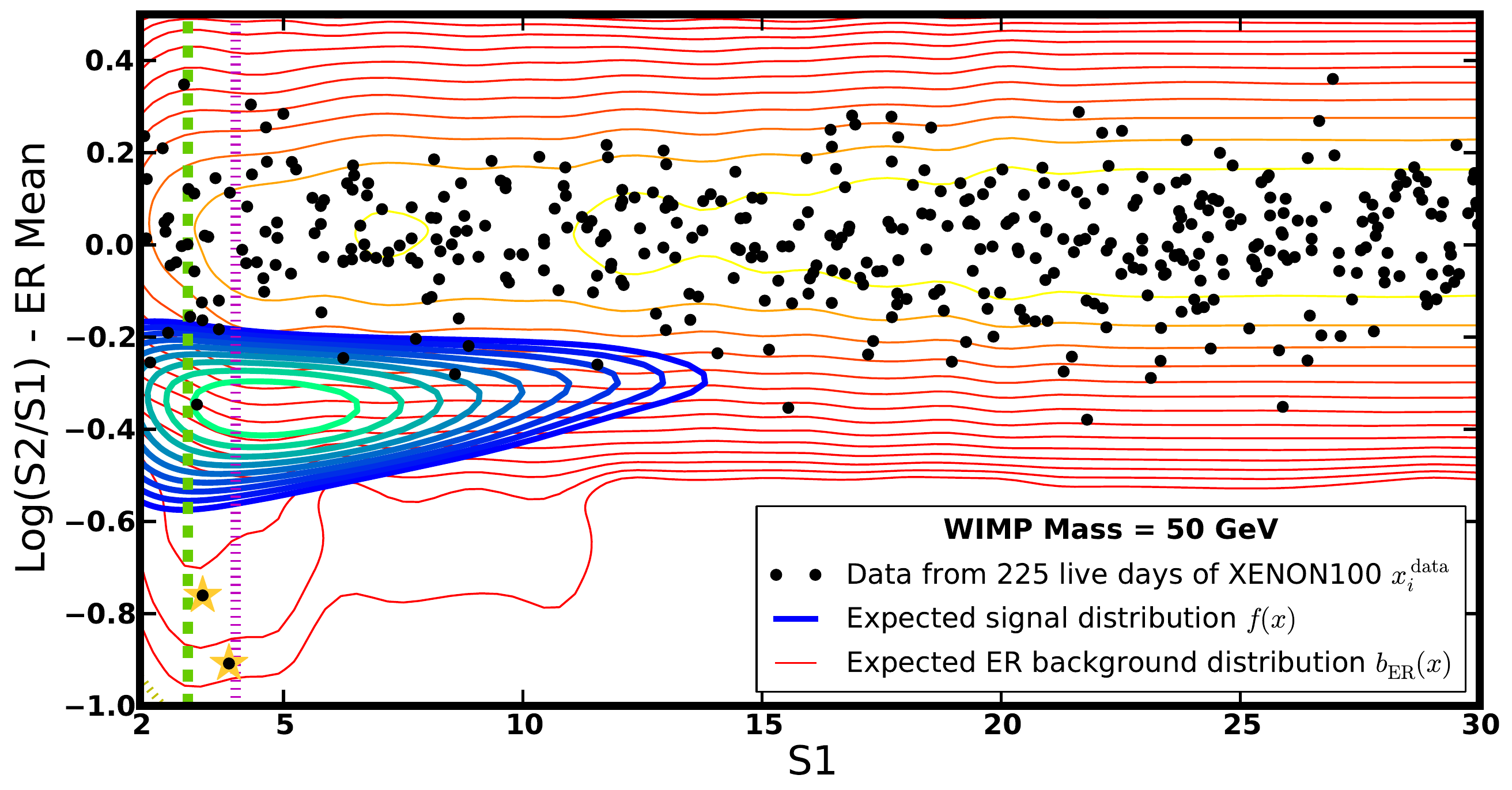}}
\subfloat{\label{fig:dists_100_b} \includegraphics[width=0.43\textwidth,trim=0 25 0 0,clip=True]{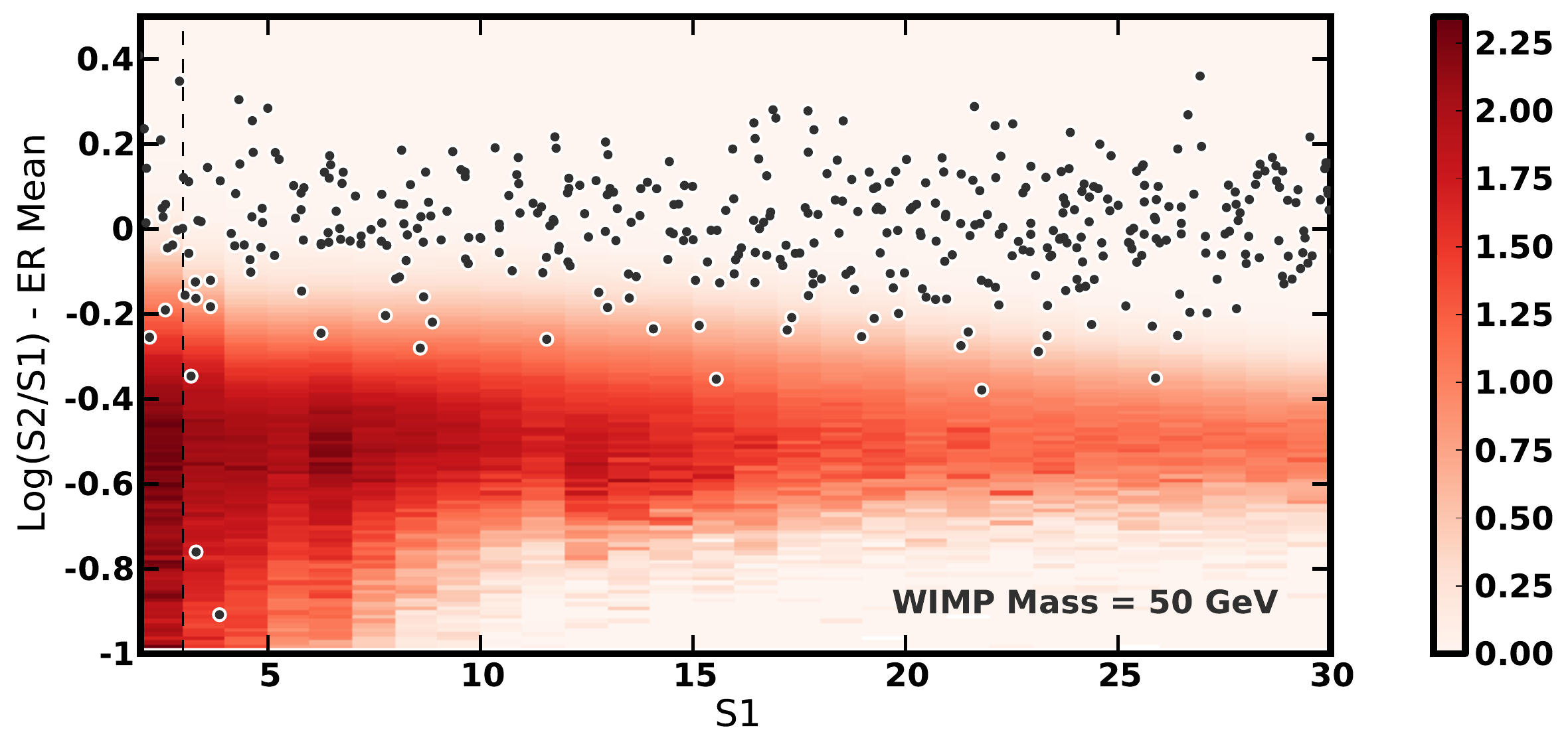}} \\
\subfloat{\label{fig:dists_100_c} \includegraphics[width=0.40\textwidth]{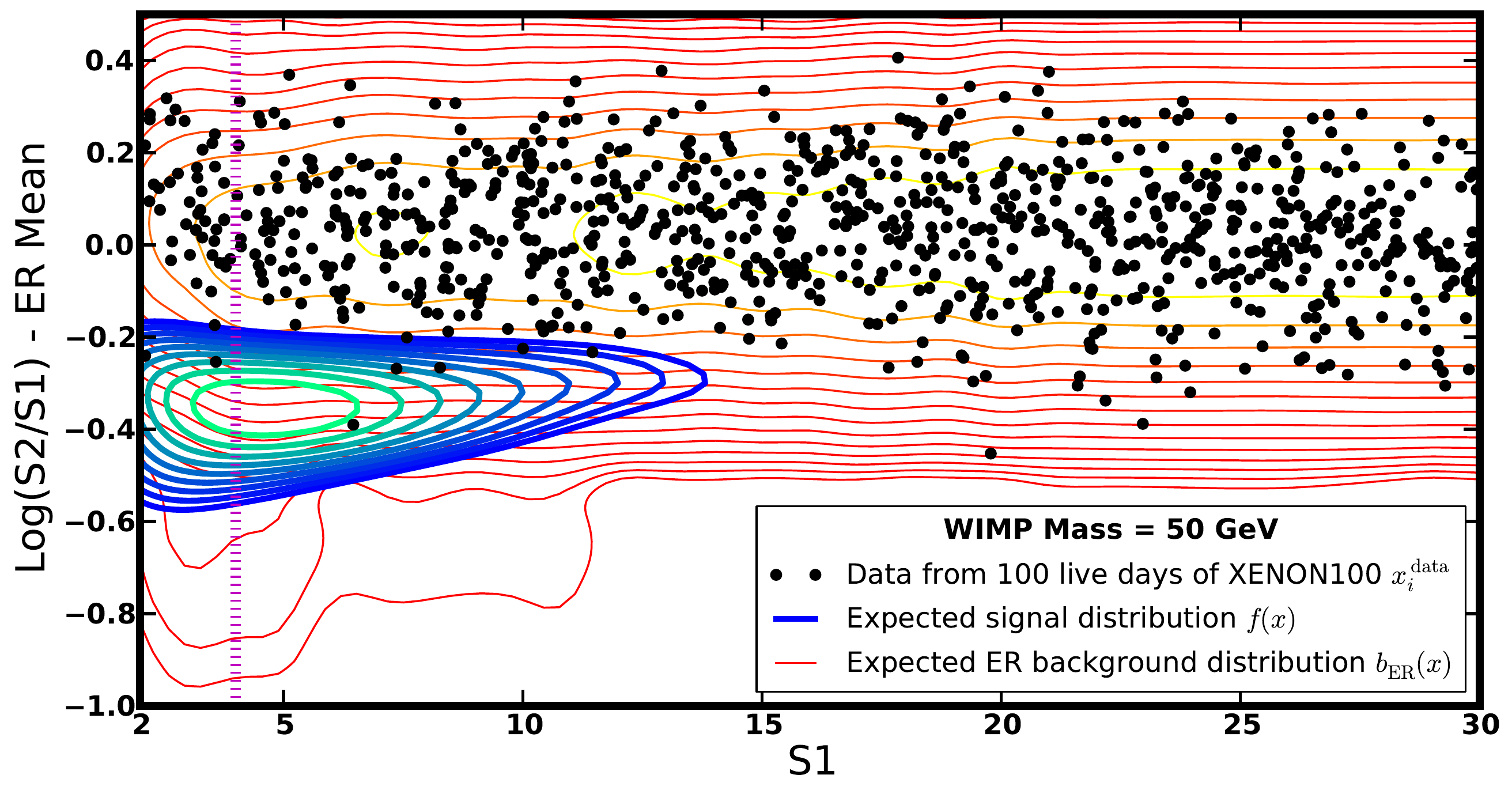}}
\subfloat{\label{fig:dists_100_d} \includegraphics[width=0.43\textwidth]{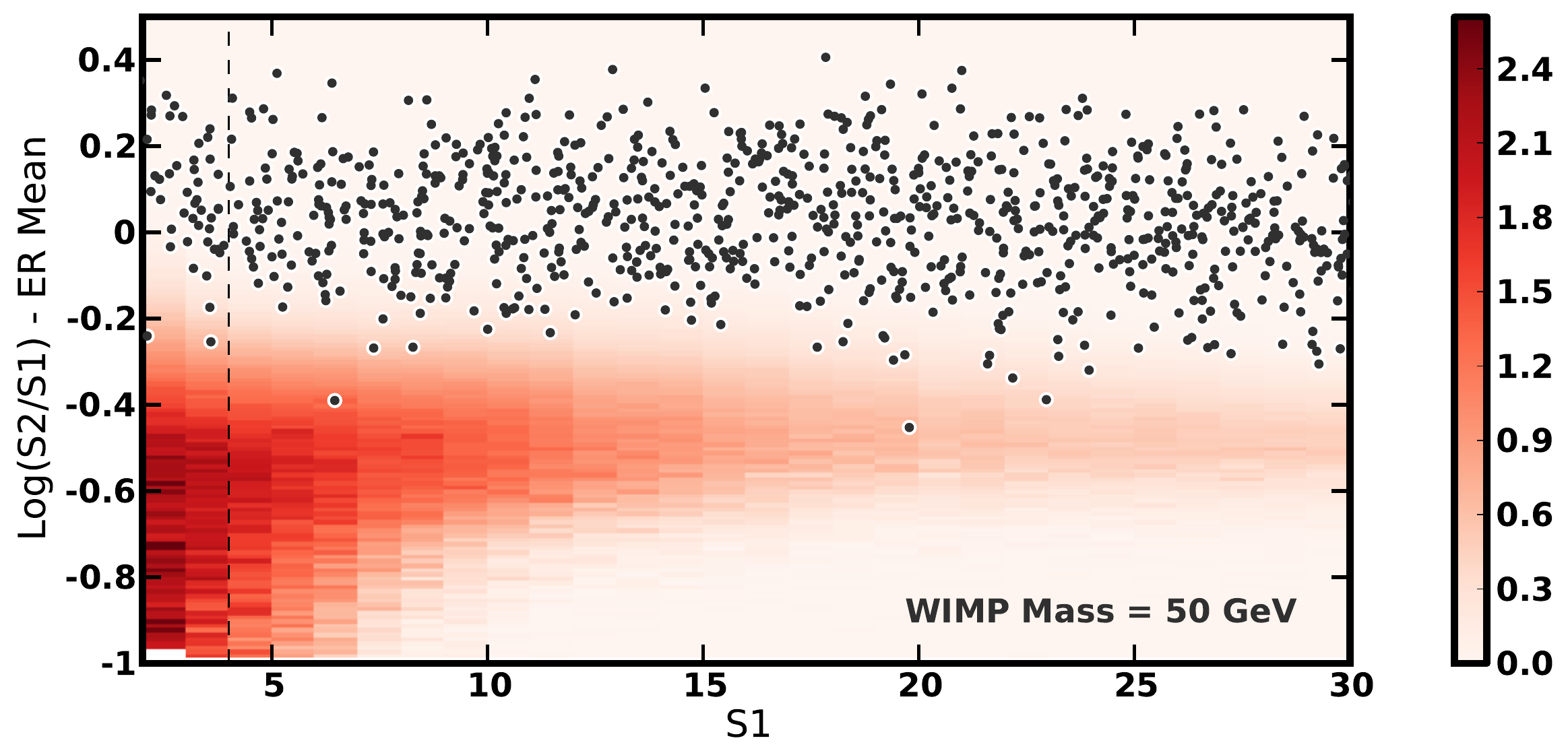}}
\caption{The upper four panels show the 225 Live Days dataset \cite{Aprile:2012_new}, while the lower two display the data for 100 Live Days of the XENON100 experiment \cite{Aprile:2011hi}. The left panels show the expected signal $f(x)$ and background $b(x)$ distributions used for our analysis. For the signal distribution, each contour is $1.2$ times less than the previous, from light to dark blue, while for the background the ratio is $1.5$ from orange to red. The data are shown as black circles. Note that only the ER background is shown here for convenience, where one can also see the anomalous background component at low-S1, as discussed in section \ref{sec:background}. For the 225 Live Days data, the two most signal-like points have been highlighted with yellow stars and are referred to as ``hint" points in the text. In the right panels we show the function $\mathrm{Ln} (1 + w(x)\cdot s)$, where $s = 10^{-8}$ here and $w(x) = f(x)/b(x)$, the weight distribution of eqn. \ref{eq:sigma_limit}. We bin $w(x)$ in units of $\Delta S1 = 0.5$ and $\Delta \mathrm{Log}(S2/S1) = 0.01 $, and interpolate between these bins for the analysis. The y-axis is shifted by the mean of the electronic-recoil band, as shown by ``ER Mean". For the 225 Live Days data we make use of two sets of cuts on the data-space: the first is to consider only points between S1$_{\mathrm{low}} = 3 \, \mathrm{PE}$ and S1$_{\mathrm{up}} = 30 \, \mathrm{PE}$, while the second moves the lower cut to S1$_{\mathrm{low}} = 4 \, \mathrm{PE}$.  The former is referred to as the full data-set in the text, while the latter removes the ``hint" points and is referred to as the ``hint"-removed data-set. For the 100 Live Days data S1$_{\mathrm{low}} = 4 \, \mathrm{PE}$ and S1$_{\mathrm{up}} = 30 \, \mathrm{PE}$. Additionally the 225 days data is bounded from below by S2 = $150 \, \mathrm{PE}$, while for 100 Live Days this moves to $300 \, \mathrm{PE}$. }
\label{fig:wimp_dists}
\end{figure*}
The strongest limits (for $m > 10 \, \mathrm{GeV}$) on the spin independent cross section for dark matter elastic scattering with nuclei have been set by xenon-based experiments i.e. XENON100 and LUX \cite{Akerib:2013tjd}. {We focus on the XENON100 experiment \cite{Aprile:2012_new,Aprile:2011hi,Aprile:2010um} as a worked example}, which operates using both liquid and gaseous xenon with a fiducial mass of $34 \, \mathrm{kg}$ (for the most recent data-set \cite{Aprile:2012_new}). The XENON100 detector identifies events by using two distinct signals \cite{Aprile:2011dd}: primary (S1) and secondary (S2) scintillation, the former of which is due to scintillation light originating from the liquid part of the detector, while the latter comes from ionised electrons, which drift to the gaseous part of the detector under an electric field. {The LUX detector \cite{Akerib:2013tjd} operates on a similar principle, but with a larger fiducial mass. The LUX collaboration also employ different cuts (e.g. a cut at S1 = 2 PE, instead of 3 PE) and potentially a different Likelihood function for their own analysis. Otherwise, the following discussion should be interesting for an understanding of the analysis of LUX data, as well as XENON100.}

In order to derive limits on the spin-independent cross section as a function of dark matter mass, the XENON100 collaboration employs a profile Likelihood approach \cite{Aprile:2012vw,Aprile:2011hx}. Such a method takes advantage of the distinct signatures in S1-S2 of electronic and nuclear recoils by splitting the data-space into a number of bands (23 in \cite{Aprile:2011hx} and 12 in \cite{Aprile:2012vw}). We can contrast this approach with our method, where the data-space is split into a grid of rectangular pixels, which are associated with a point in the data-space $x = (\alpha,\beta)$. Hence, we expect our gridded approach to perform better than this method of bands used by the XENON100 collaboration, since we can exploit the difference between signal and background to the maximum amount, while they are limited by the rather coarse-grained resolution of their bands\footnote{The Likelihood used by the XENON100 collaboration is claimed to be able to exploit the spectra of events within each band using a separate term in the Likelihood \cite{Aprile:2012vw,Aprile:2011hx}. However in practice it is not clear how effective this actually is, and we do not believe it exploits the difference between signal and background as well as our gridded method.}. This application should serve as a clear demonstration of the advantages to \emph{any} Direct Detection experiment of using our method.

We can identify S1 and S2 with our discrimination parameters $\alpha$ and $\beta$ from section \ref{sec:method}, though here we choose instead to take $\alpha = \mathrm{S1}$, $\beta = \mathrm{Log}(\mathrm{S2}/\mathrm{S1})$, to match more closely the method used by the XENON100 collaboration themselves (and also the LUX collaboration \cite{Akerib:2013tjd}). We will proceed first to discuss the determination of the signal $f(x)$ and background $b(x)$ distributions for the XENON100 experiment, before applying our method to data, both the more recent 225 live days data (225LD) \cite{Aprile:2012_new} and the older 100 live days data-set (100LD) \cite{Aprile:2011hi}.

\subsection{Signal Distribution}

 \subsubsection{WIMP Recoil Spectrum}
\label{sec:rec_spec}
Potential WIMP events are characterised by their recoil spectra $\frac{\mathrm{d}R}{\mathrm{d}E}$, parameterised as \cite{Lewin199687,Cerdeno:2010jj},
\begin{equation}
\label{eqn:recoil_rate}
\frac{\mathrm{d}R}{\mathrm{d}E} = \frac{\sigma (E)}{2 m \mu^2} \rho \eta (E) ,
\end{equation}
where $\sigma(E)$ is the WIMP-nucleus cross-section as a function of energy $E$, $\mu$ is the WIMP-nucleus reduced mass, $\rho = 0.3 \, \mathrm{GeVcm}^{-3}$ is the local dark matter density and $\eta(E) = \int_{v_{\mathrm{min}}}^{\infty} \mathrm{d}^3 v \frac{f(v + u_{\mathrm{e}})}{v} $ is the WIMP mean velocity. The mean velocity is integrated over the distribution of WIMP velocities in the galaxy $f(v)$ boosted into the reference frame of the Earth by $u_{\mathrm{e}}$. The lower limit of the integration is $v_{\mathrm{min}} (E)$, which is the minimum WIMP velocity required to induce a recoil of energy $E$. We assume the standard halo model such that $f(v)$ is given by a Maxwell-Boltzmann distribution cut off at an escape velocity of $v_{\mathrm{esc}} = 544 \, \mathrm{kms}^{-1}$.

We assume that WIMPs interact identically with protons and neutrons\footnote{We take a simple model of DM-nucleon elastic scattering here for convenience, however our method is easily generalised to more complicated models (e.g. \cite{DelNobile:2012tx,Foot:2013msa,Schwetz:2011xm,Chang:2010yk,Hooper:2011hd}) by replacing $\sigma(E)$ and $\mathrm{d}R / \mathrm{d} E$.} giving $\sigma(E) = \sigma \left( \frac{\mu}{\mu_p} \right)^2 A^2 F^2(E)$, where $\sigma$ is the zero-momentum WIMP-nucleon cross section, $A$ is the atomic mass of xenon, $\mu_p$ is the WIMP-proton reduced mass and $F(E)$ is the Helm nuclear form factor \cite{Lewin199687}. For 225LD (100LD) we use a value of $224.6$ days ($100.9$ days) for the exposure and $34$ kg ($48$ kg) for the target mass.

\subsubsection{Calculation of S1 and S2 for Nuclear-Recoils}

At a given nuclear-recoil energy $E$ the expected primary ($S1_{\mathrm{exp}}$) and secondary ($S2_{\mathrm{exp}}$) scintillation signals are obtained from the following formulae \cite{Aprile:2012vw,Sorensen:2010hq,Savage:2010tg,Bezrukov:2010qa},
\begin{eqnarray}
S1_{\mathrm{exp}} &=& P \left( E \cdot L_{\mathrm{eff}}(E) \cdot L_y \cdot \frac{S_n}{S_e} \right) \label{eqn:S1} \\
S2_{\mathrm{exp}} &=& Y \cdot P \left(E \cdot Q_y(E) \right) \label{eqn:S2},
\end{eqnarray}
where $P(x)$ represents a Poisson distribution with expectation value $x$, $L_y = 2.20 \pm 0.09 \, \mathrm{PE} \, \mathrm{keV}^{-1}$, $ \frac{S_n}{S_e} = 0.95/0.58$, $Y$ is a gaussian-distributed value with mean $19.5$ PE per electron and width $\sigma = 6.7 \, \mathrm{PE}/e^-$ \cite{Aprile:2013teh} , $L_{\mathrm{eff}} (E)$ is the relative scintillation efficiency and $Q_y (E)$ is the ionisation yield. For $Q_y$ there is a degree of uncertainty on its functional form \cite{Aprile:2013teh}; we use the model of \cite{Bezrukov:2010qa} in this work, however we have obtained similar results with the best-fit curve from \cite{Aprile:2013teh}. $L_{\mathrm{eff}}$ is obtained from a cubic spline fit to data from \cite{Chepel:2006yv,Manzur:2009hp,Plante:2011hw,Aprile:2008rc}.

To obtain the $S1_{\mathrm{obs}}$ and $S2_{\mathrm{obs}}$ signals observed in the detector, we must include the finite detector resolution and the cuts imposed by the XENON100 collaboration on the data \cite{Aprile:2013teh,Aprile:2012vw,Aprile:2011hi}. Both $S1_{\mathrm{exp}}$ and $S2_{\mathrm{exp}}$ are blurred with a gaussian of width $0.5 \sqrt{n}$ for $n$ photoelectrons (PE) to take account of the finite photomultiplier (PMT) resolution \cite{Aprile:2011hx}. The effect of cuts is then implemented using the cut-acceptance curve as a function of S1 \cite{Aprile:2013teh,Aprile:2012vw} after applying the resolution effect. Additionally an S2 threshold cut is applied before gaussian blurring, cutting away all points with $\mathrm{S1} < 1 \, \mathrm{PE}$ \cite{Aprile:2012_new}.

\subsubsection{Expected Dark Matter signal in XENON100} 

The expected signal distribution for a given WIMP mass in the data-space $f(x)$ can now be calculated using $\mathrm{d}R/\mathrm{d}E$ of section \ref{sec:rec_spec}, at a value of the reference cross-section $\sigma_0 = 10^{-35}  \, \mathrm{cm}^2$ (or $10^{-34}  \, \mathrm{cm}^2$ for $m < 10 \, \mathrm{GeV}$). The energy range between $1 \, \mathrm{keV}$ and $60 \, \mathrm{keV}$ is separated into bins of size $\Delta E = 0.01 \, \mathrm{keV}$. For each binned energy $E_{\mathrm{rec}}$ we calculate $S1_{\mathrm{obs}}$ and $S2_{\mathrm{obs}}$ a total of $N_{\mathrm{rec}}$ times, where $N_{\mathrm{rec}} = \frac{\mathrm{d}R}{\mathrm{d}E} (\sigma_0, E_{\mathrm{rec}}) \Delta E$, to obtain the full signal distribution as expected in XENON100. The result is shown for two different masses in fig. \ref{fig:wimp_dists}. Similar simulations of the signal distribution expected from XENON100 have been performed in \cite{Sorensen:2012ts,Hooper:2013cwa,Aprile:2013teh}, however our method goes further and directly links these to the analysis through the weight function $w(x) = f(x) / b(x)$, as shown in figure \ref{fig:wimp_dists}.

\subsection{Background Distribution in XENON100 \label{sec:background}}

The expected distribution of electronic-recoil background events $b_{\mathrm{ER}}(x)$ is determined from fits to $^{60}$Co calibration data\footnote{Our determination of  $b_{\mathrm{ER}}(x)$ would improve were we to use the $^{232}$Th calibration data (especially for the anomalous component), collected by the XENON100 collaboration for their most recent analysis \cite{Aprile:2012_new}, however this is not currently publicly available.}, as is done in \cite{Aprile:2011hi,Aprile:2011hx}. Although the electronic recoil events appear mostly Gaussian distributed,  the XENON100 collaboration noticed the presence of an anomalous (non-Gaussian) background component \cite{Aprile:2011hi}. This could be due to double-scatter gamma events, where only one of the gammas contributes to the S2 signal. Both such components of the ER background are included, indeed the anomalous component can be seen in figure \ref{fig:wimp_dists} predominantly at low-S1. The distribution is normalised by the total number of expected background events, whose rate takes the constant value of \textcolor{black}{$0.0053$ counts per day per kg per keV$_{\mathrm{ee}}$ \cite{Aprile:2011vb,Aprile:2012_new}. For 100LD the background is larger due to krypton contamination in the experimental apparatus, taking a value of $0.022$ counts per day per kg per keV$_{\mathrm{ee}}$ \cite{Aprile:2011hi}}.

We also model the nuclear-recoil background due to neutrons $b_{\mathrm{NR}}(x)$. The distribution is calculated as for the signal distribution, but replacing $\mathrm{d}R/\mathrm{d}E$ with the expected energy spectrum of neutron scatters in the detector \cite{Aprile:2013tov}. Hence the total background distribution is $b(x) = b_{\mathrm{ER}}(x) + b_{\mathrm{NR}}(x)$.

\begin{figure*}[t]
\centering
\subfloat{\includegraphics[scale=0.56,trim=0 10 20 25,clip=True]{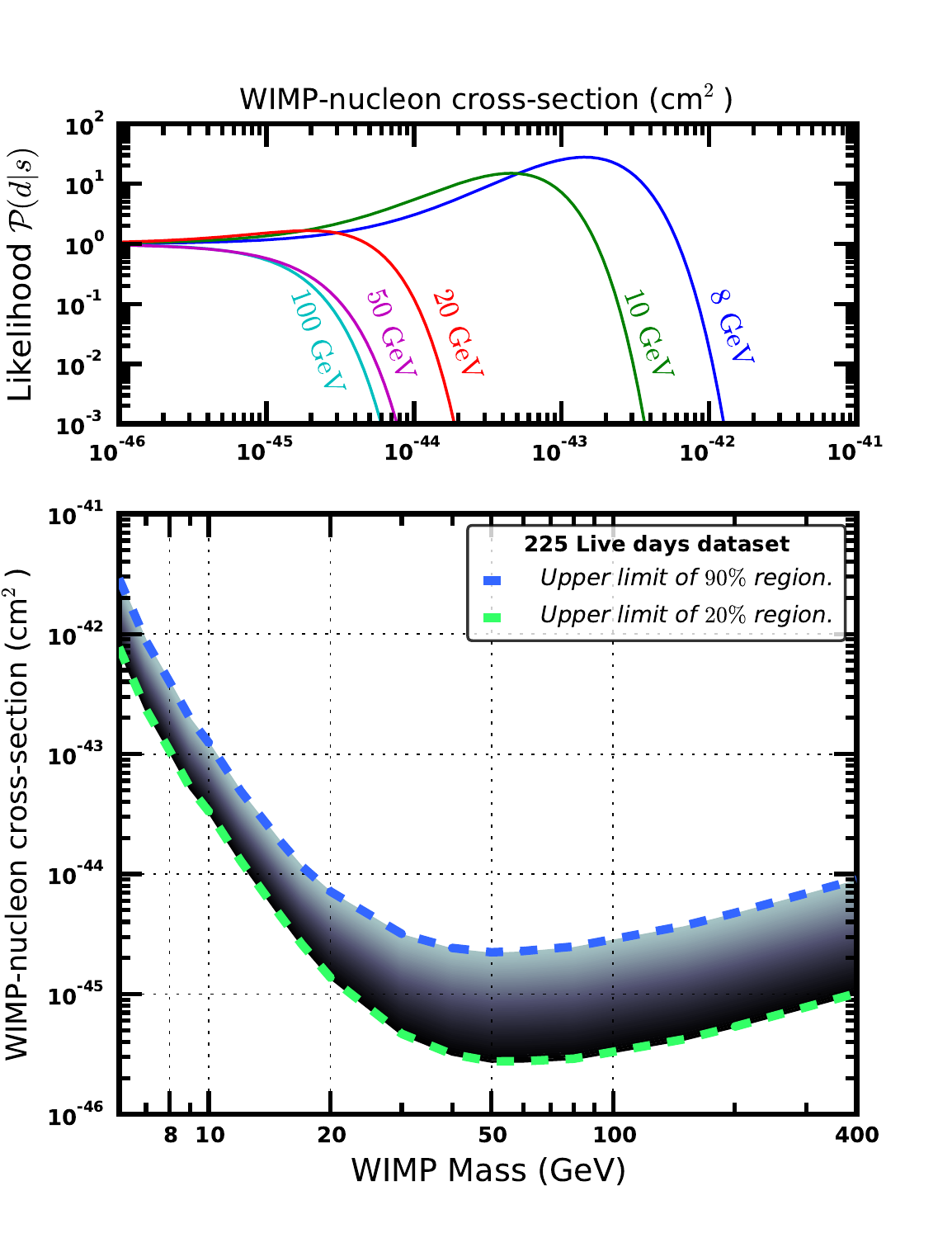}} 
\subfloat{\includegraphics[scale=0.56,trim=13.5 10 20 25,clip=True]{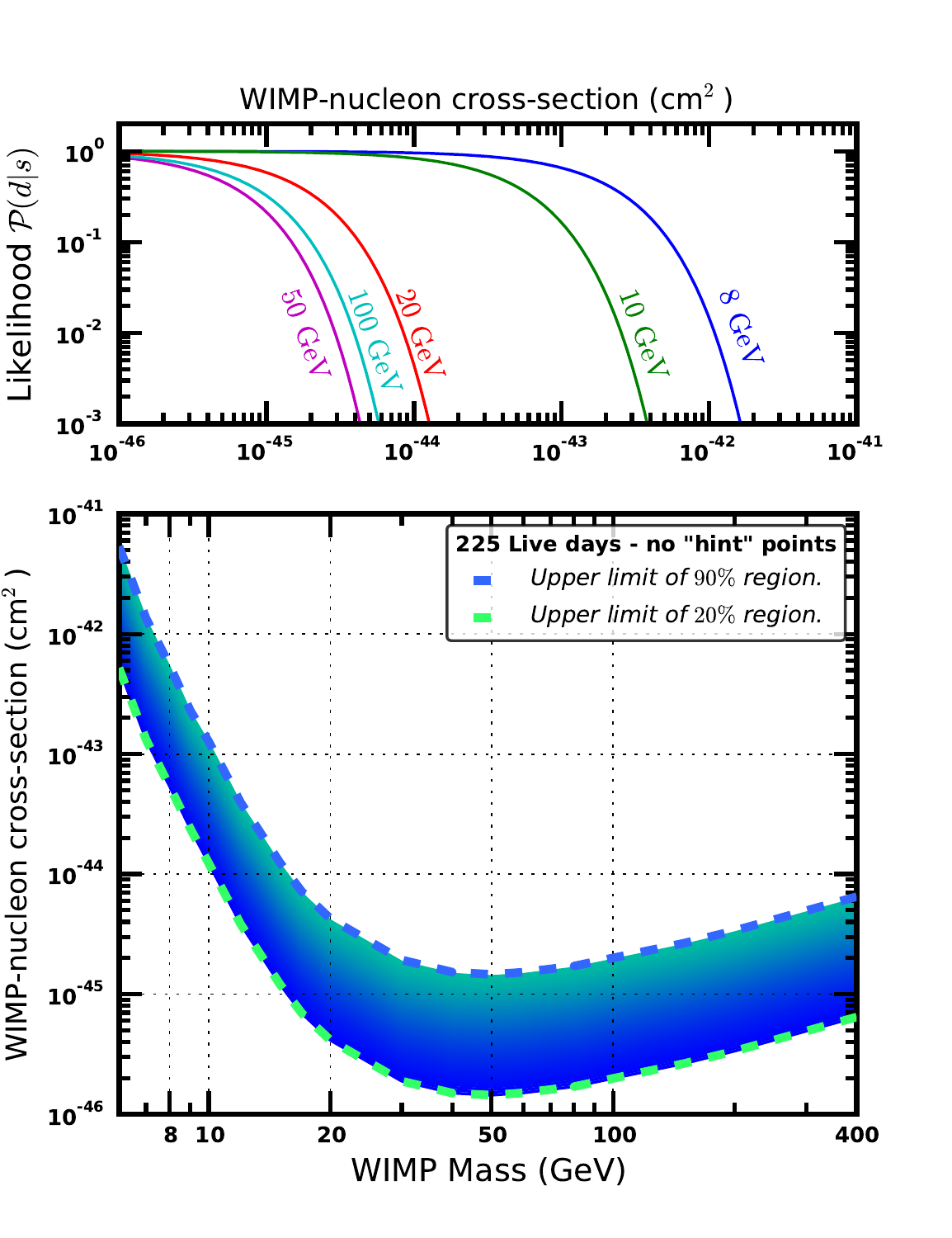}} 
\subfloat{\includegraphics[scale=0.56,trim=13.5 10 20 25,clip=True]{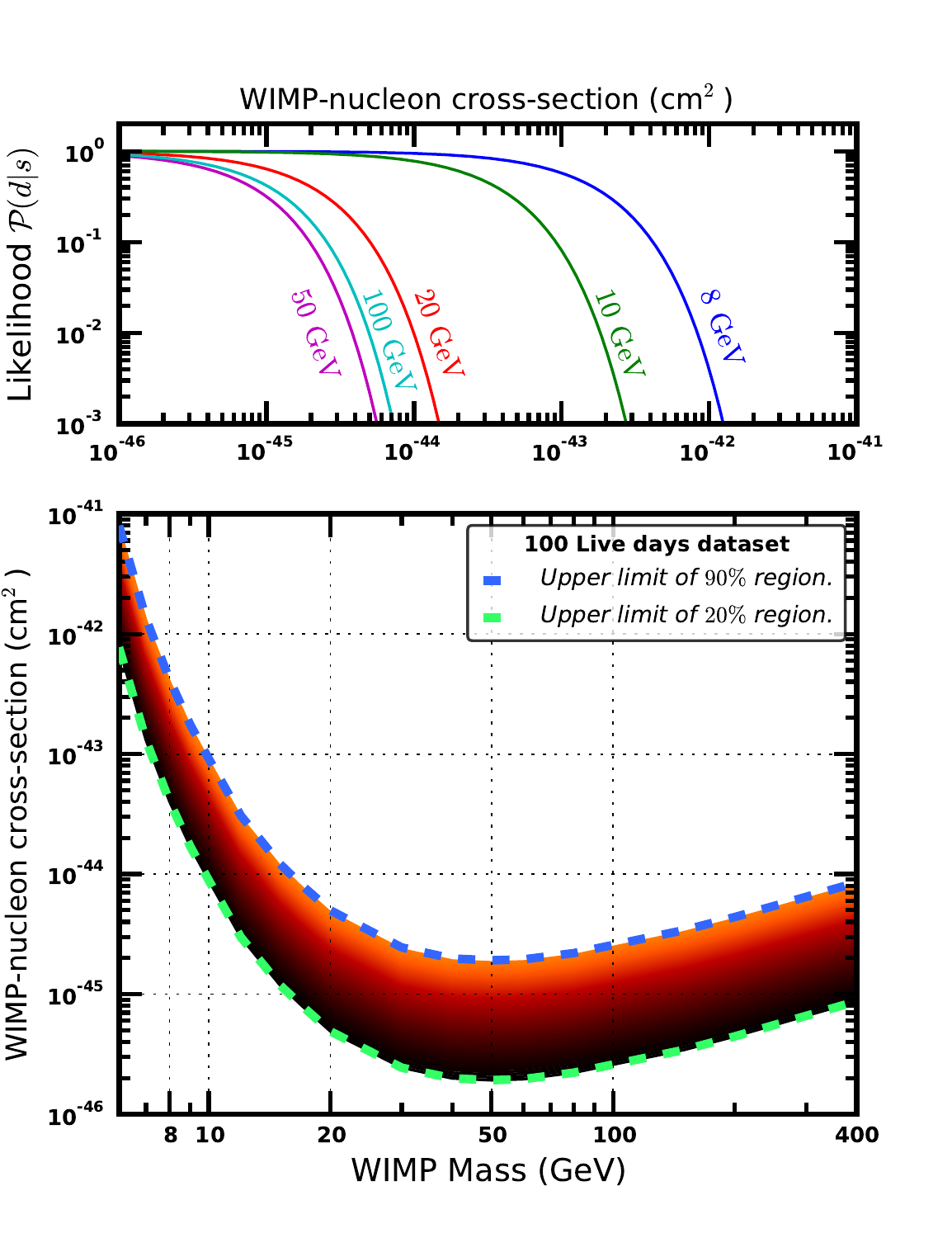}} 
\caption{Plots showing exclusion limits and regions of credibility, derived from applying our analysis to data from the XENON100 experiment \cite{Aprile:2012_new}. For the left-most 225LD analysis, there is a weak preference for low-mass DM, which vanishes under more stringent cuts (central) or with the 100LD data (right). The upper panels show examples of the (un-normalised) Likelihood function $\mathcal{P}(d|s)$ for various WIMP masses, while the lower panels show the result of integrating the posterior from $s = 0$ up to some limiting value, in order to define an exclusion limit for a given significance. The region between the two dashed lines shows exclusion curves with significance increasing linearly from darker to lighter shading. One can indeed consider this region as one of $70 \%$ significance. For the left panels we have used the full 225LD dataset (all points between S1$_{\mathrm{low}} = 3 \, \mathrm{PE}$ and S1$_{\mathrm{up}} = 30 \, \mathrm{PE}$), while for the central panels the analysis has been performed with the two most signal-like (labelled as ``hint") data-points removed by cutting off the data-space below S1$_{\mathrm{low}} = 4 \, \mathrm{PE}$. The right-most panels show results for the 100LD data.}
\label{fig:lims}
\end{figure*}

\subsection{Example Exclusion Limits and Posterior Scans}
\subsubsection{Signals and Limits from XENON100 Data}
Now that we know how to calculate the expected signal and background distributions $f(x)$ and $b(x)$, we are ready to apply our method to the data from the XENON100 experiment. All relevant ingredients are displayed in fig. \ref{fig:wimp_dists}; the left panels show the regions where the expected signal and background are expected to be largest, while the right panels show plots of $\mathrm{Ln} (1 + w(x) \cdot s)$ as used directly for our analysis. The discrimination between signal and background is maximised provided the two-dimensional bins for $w(x) = f(x) / b(x)$ are small enough: data-points where $w(x)$ is large are more likely to be due to signal than background, while the opposite is true for points located where $w(x)$ is small. This is then fed directly into our analysis, hence figure \ref{fig:wimp_dists} contains all of the main ingredients of our method.

Shown in figure \ref{fig:lims} are the results of applying our method to the data. In order to understand the effect of data-points consistent with a signal interperetation, we have performed the analysis with both the full dataset (with a lower cut on S1 at $S1_{\mathrm{low}} = 3 \, \mathrm{PE}$), and with a reduced dataset, where the two ``hint" data-points (i.e. the starred points in figure \ref{fig:wimp_dists} ) have been removed by cutting away the data-space below $S1_{\mathrm{low}} = 4 \, \mathrm{PE}$ \footnote{We could instead have moved the low-S2 cut from $150 \, \mathrm{PE}$ to $300 \, \mathrm{PE}$, as for the 100LD data-set, which would remove one of these points.}. The former is displayed in the left panel of fig. \ref{fig:lims}, while the results for the reduced dataset are shown on the central panel. Results from the 100LD data are shown on the right.

As discussed in section \ref{sec:sigs_lims} we can define regions of credibility (either exclusion limits or potential discovery regions) by integrating under the normalised posterior $\mathcal{P}(s|d)$. Hence in the lower panels of figure \ref{fig:lims} we show exclusion limits for various levels of confidence, between $20 \%$ and $90 \%$, calculated by integrating the posterior from $s = 0$ up to the limiting value of $s$. One can equivalently consider the parameter space between these limits as a region of $70 \%$ credibility. {The $90 \%$ limit for the full 225LD data-set can be compared with the result from \cite{Aprile:2012_new}, while the shaded band represents how the limit changes with different confidence.}

The upper panels show the dependence of the Likelihood $\mathcal{P}(d|s)$ as a function of $\sigma$ for various WIMP masses. One can see directly that for the full 225LD dataset the Likelihood function has a maximum (corresponding to a minimum in the Hamiltonian), indicating a preference for the data of a particular value of $\sigma$, which is strongest for lighter WIMPs. Indeed this can also be observed in the exclusion curve as we change the significance value: particularly for lighter WIMPs the region of credibility between the $20 \%$ and $90 \%$ limits is denser as compared to heavier DM. This is due directly to the presence of a maximum in the Posterior and Likelihood. 

This is particularly interesting in the context of the potential hints of light DM in CDMS \cite{Agnese:2013rvf} and CoGeNT \cite{Aalseth:2011wp} (and to some extent DAMA). However the significance of such a hint is weak. Indeed the $50 \%$ credible region for an $8 \, \mathrm{GeV}$ WIMP lies between $6.15 \cdot 10^{-44} \, \mathrm{cm}^2$ and $2.15 \cdot 10^{-43} \, \mathrm{cm}^2$, with a best-fit cross section at $1.40 \cdot 10^{-43} \, \mathrm{cm}^2$. Of course the cross-section is still inconsistent with the best-fit region from CDMS \cite{Agnese:2013rvf}, unless one changes the systematic parameters to a rather extreme degree \cite{Hooper:2013cwa} or considers less standard interactions \cite{Frandsen:2013cna}. 

Claims that these points are consistent with a DM signal are likely to be overly optimistic. The significance of the signal is comparable to a $1 \sigma$ fluctuation\footnote{Since our method is Bayesian, a comparison with frequentist confidence intervals is not directly possible. However if one considers a $1 \sigma$ confidence interval as (roughly) comparable to a $68 \%$ region of credibility, then we actually find the significance to be a bit less than $1 \sigma$. Indeed our choice of $50 \%$ was motivated by the fact that it is close to the largest two-sided interval we could set around the maximum-likelihood value of cross section. The sigma-level is only approximate though, as our Likelihood is non-gaussian (see fig. \ref{fig:lims}).}, and hence these data-points may just be events from the non-gaussian ER background, which we already model. We can additionally compute the Bayes factor $\mathcal{B}$ \cite{doi:10.1080/01621459.1995.10476572} for e.g. an $8 \, \mathrm{GeV}$ WIMP, by calculating the ratio of the joint signal and data probability $\mathcal{P}(d,s)$ integrated over all $r$, to $\mathcal{P}(d,s (r = 0))$ i.e. the no-DM scenario, where $\sigma = 0$. Hence the size of $\mathcal{B}$ should tell us to what degree a positive signal of DM is preferred, relative to the scenario where no signal is present (see \cite{doi:10.1080/01621459.1995.10476572} for details). We calculate $\mathcal{B} = 3.18$, which is just on the boundary of being a positive result. Hence, again we can conclude there is only a weak hint of signal for a low-mass WIMP. There are also systematic uncertainties from $L_{\mathrm{eff}}$ and $Q_y$, though they are unlikely to result in a significant enhancement of the signal significance.

Indeed, as can be seen from \ref{fig:wimp_dists} if one attributes these points to a WIMP signal, one must also explain why no data is seen where the signal from DM is expected to be even larger, at lower values of S1 for example. Even so, the presence of consistency with signal, however weak, indicates some sort of new phenomenon may be present: either DM or an unknown (or possibly misunderstood)  background. Hence an interpretation of these points in terms of Dark Matter is possible but premature, however they are instructive as an example of the effect of signal-like points on our ability to set limits on light DM.

By contrast when the two ``hint" data-points are removed from the analysis by the more stringent low-S1 cut (see figure \ref{fig:wimp_dists} for details), there is no maximum in the Likelihood and Posterior for any WIMP mass, as one would expect since all points are in a region where the weight $w(x) = f(x)/b(x)$ is small. Indeed the density of the posterior is now less for all masses than for the full data-set, with the contrast particularly stark for lighter DM. The same is seen for 100LD, for which no hint of signal is present. In addition, the limits without the ``hint" points are stronger since the data are now almost completely consistent with a negative result. If the XENON100 collaboration were to observe additional signal-like points in their data, one would expect the density of the posterior to increase around the best-fit region.

In any case this demonstrates the ability of our method to accurately set limits or define potential discovery regions. All of the relevant information is contained within the posterior $\mathcal{P}(s|d)$, which can be integrated over to define the degree of belief that a given region of parameter space is consistent with the data.

\subsubsection{Comparison with results from XENON100}
\begin{figure}[htb]
\centering
\includegraphics[scale=0.54,trim=0 10 10 30,clip=True]{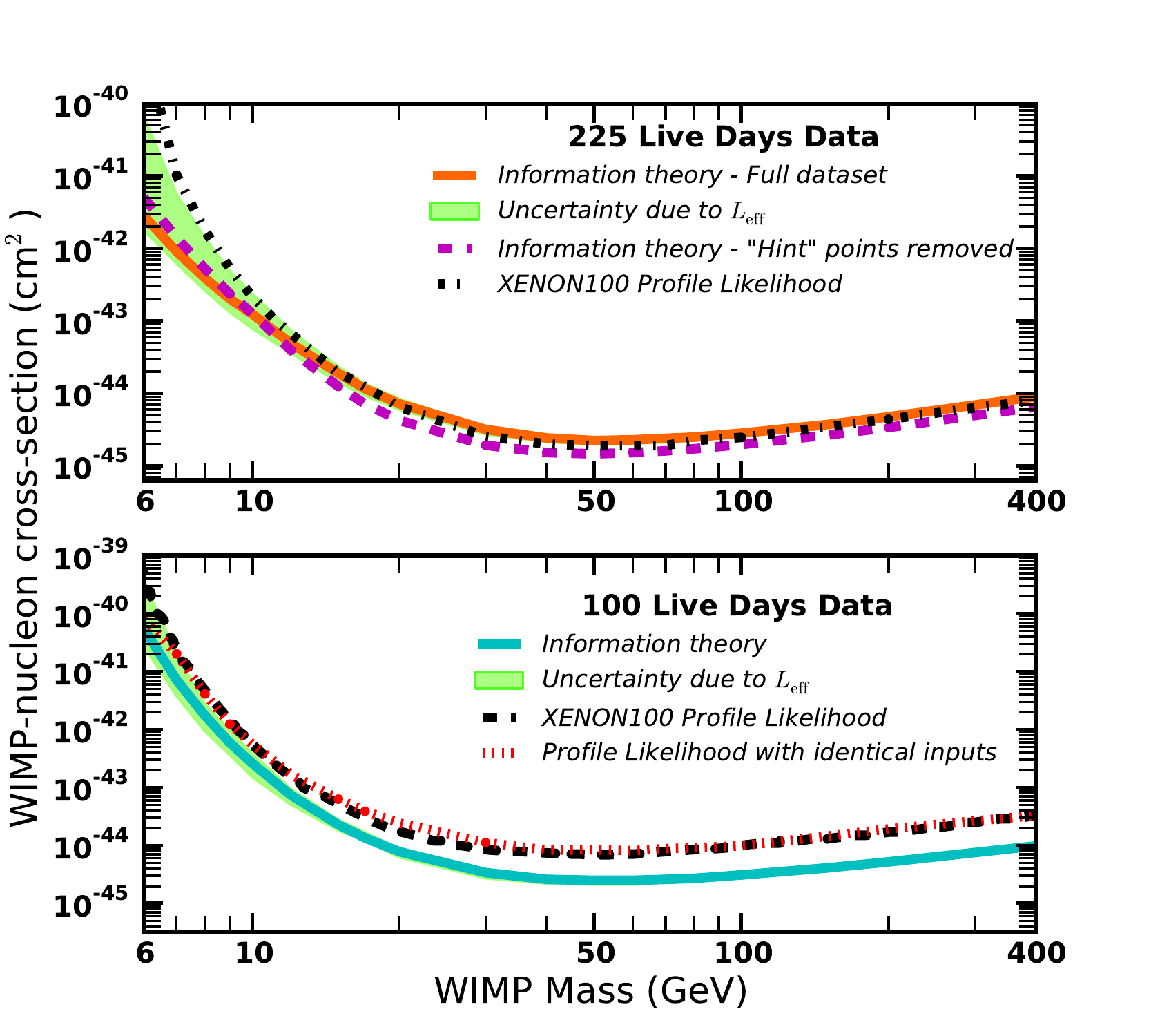}
\caption{A comparison of various limits set with either 225 live days \cite{Aprile:2012_new} or 100 live days \cite{Aprile:2011hi} of XENON100 data. Limits from information theory refer to those derived using the method presented in this work. For the 100 Live Days data we also compare the result of a profile Likelihood analysis performed by the XENON100 collaboration with that from an analysis we have done using the same profile Likelihood method, but where the inputs are identical to those for our Bayesian method, such as $f_{\mathrm{NR}}(x)$ and $b(x)$. The limit from our Bayesian information theory method agrees with the XENON100 published limit for 225LD, but is several times stronger for 100LD.}
\label{fig:limit_comparison}
\end{figure}

Before forming any firm conclusions on the efficacy of our method in searching for Dark Matter signals in Direct Detection data, we must compare our results to those previously found by the XENON100 collaboration. Shown in figure \ref{fig:limit_comparison} is our $90 \%$ confidence limit (identical to the one in figure \ref{fig:lims}), compared with the limit derived by the XENON100 collaboration with the same 225 live days dataset \cite{Aprile:2012_new}, but their own profile Likelihood analysis \cite{Aprile:2011hx}. Uncertainties due to the relative scintillation efficiency $L_{\mathrm{eff}}$ are shown as a shaded region around our limit (see e.g. \cite{Savage:2010tg,Davis:2012vy} for a review).

 In addition, in the lower panel of figure \ref{fig:limit_comparison} we also show the results of applying our method to the 100 live days dataset, along with the limit from the XENON100 collaboration using their profile Likelihood method, and a limit we have independently derived using the same method, but with identical inputs to our information theory analysis.

The exclusion limit derived with our information Hamiltonian method agrees with that derived by the XENON100 collaboration for the 225 live days data-set for large masses. For lighter WIMPs our limit is stronger, though this is likely due to uncertainty in the low-energy extrapolation of $L_{\mathrm{eff}}$ \cite{Davis:2012vy}. Indeed the XENON100 collaboration employ the most conservative approach and cut $L_{\mathrm{eff}}$ to zero below $3 \, \mathrm{keV}_{\mathrm{nr}}$, where no data is available \cite{Aprile:2012_new}. Our limit is derived using a constant extrapolation instead, though the uncertainty band shows the limit under different parameterisations of $L_{\mathrm{eff}}$\cite{Davis:2012vy}. Hence one can consider our result as an independent cross-check of the limit published by the XENON100 collaboration.

There are undoubtably other small differences between our inputs and those used by the XENON100 collaboration, however the agreement of both limits indicates that our method does indeed perform correctly when analysing Direct Detection data. Note also that for the ``hint"-removed data-set, where the low-S1 cut is moved to S1$_{\mathrm{low}} = 4 \, \mathrm{PE}$, the limit is stronger for heavy WIMPs due to the removal of the signal-like points by the cut. This is not so for lighter WIMPs, since much of the region where one expects to see signal is cut away in addition to the ``hint" points.

We note however that when applying our method to the 100LD data \cite{Aprile:2011hi} that our information theory limit is stronger than that derived using the profile Likelihood analysis, both performed directly by the XENON100 collaboration and from an independent analysis we have carried out. Since the latter two limits are in agreement, it would be difficult to blame the inputs of the analysis on this discrepancy between the limits, hence it is likely that the coarse-graining\footnote{Specifically we refer to the splitting of the data-space into a finite number of bands for the profile Likelihood method used by the XENON100 collaboration, which necessarily limits the amount of information extracted from the data, as opposed to our method where the data-space is pixelated (see figure \ref{fig:wimp_dists}). } of the profile Likelihood analysis has resulted in the derivation of an over-conservative limit. {To reiterate: we refer specifically to the profile Likelihood analysis used by XENON100 here. The issue is not with the frequentist method itself, but rather with the choice of Likelihood function used by the collaboration. Hence, our limit is more accurate because we use a Likelihood which exploits the whole data-space, and this should also be reflected in a profile Likelihood analysis which followed the same principles.}

The reason for this discrepancy arising only for the 100LD dataset is not entirely clear, though it is likely that the increased background in this dataset relative to that from 225 live days \cite{Aprile:2012_new} (due to the krypton leakage) has effectively fooled the analysis into treating too many points as potential signal, thereby weakening the limit. Hence we believe that this demonstrates the robustness of our method as compared to such a profile Likelihood analysis, since it is less susceptible to leakage of background points into the signal region.

\section{Conclusion}
\label{sec:conc}

In this work we have introduced a Bayesian method of analysing data from Dark Matter Direct Detection experiments. Our method takes as input the data itself and the expected signal and background distributions, defined over the whole data-space, which is divided into a grid of two-dimensional pixels. This enables us to take full advantage of the distinct expected distributions signal and background events, and hence to set limits (or discovery regions) without resorting to conservative approximations.

Using data from the XENON100 experiment \cite{Aprile:2012_new} as a worked example we demonstrated how one would apply our method to Direct Detection data. {This has direct relevance also to LUX experiment \cite{Akerib:2013tjd}, and any future runs of XENON100.} We have shown that there is merit in looking beyond the $90 \%$ confidence limit, as hints of signal may be affecting the structure of the Likelihood and Posterior in a non-trivial manner. Indeed an analysis of the XENON100 data from 225 Live Days indicates a weak preference in the data  for a light DM particle. At 50$\%$ confidence the best fit cross section is in between $6.15 \cdot 10^{-44} \, \mathrm{cm}^2$ and $2.15 \cdot 10^{-43} \, \mathrm{cm}^2$ for an $8 \, \mathrm{GeV}$ WIMP; the error bars being relatively large, it is very premature to argue that this is evidence for Dark Matter. Similar regions can be obtained for any dark matter particle with a mass below $\sim 20$ GeV, with a possible evidence for a dark matter signal in the data vanishing for masses above about 20 GeV. If indeed these points are due to a detection of Dark Matter, more data from the XENON100 experiment should increase the confidence level and shrink the error bars on the cross section. Alternatively, these events may be found to be due to an additional background process or the anomalous component of the ER background, in which case the signal significance would vanish with more data. {Considering the recent null result from the LUX experiment \cite{Akerib:2013tjd}, the latter would seem to be a more plausible explanation.}

We also demonstrated that our new method can produce a complementary analysis to the one currently used by the XENON100 collaboration, where the data are placed into bands.  Indeed our limit and theirs agree for the most recent 225 Live Days data-set \cite{Aprile:2012_new}, however ours is several times stronger for the data from 100 Live Days \cite{Aprile:2011hi}. The reason for this disagreement for the older data-set is not clear. However it is possible that since the background was higher due to krypton contamination, there was a greater proportion of background events leaking into the region where signal was expected (i.e. the more signal-like bands of the analysis used by the XENON100 collaboration), which may have fooled their analysis into setting too weak a limit.  Additionally our method could be even more robust, especially if one exploits the full detector volume (with $f$ and $b$ now depending on physical positions in the detector).

Our analysis can be seen as an independent analysis of the XENON100 data, and more importantly could  be employed by any present or forthcoming experimental collaboration for such a purpose. {In particular, our method can be easily applied to the LUX experiment \cite{Akerib:2013tjd}, since it operates on a similar principle to XENON100. In this case one should hope to find agreement with our Bayesian results and the frequentist method used by the LUX collaboration, which should provide an important cross-check of the LUX results. Future experiments such as XENON1T \cite{Aprile:2012zx}, LZ \cite{Malling:2011va} and SuperCDMS \cite{Agnese:2013jaa} could also benefit from a Bayesian cross-check.}

The use of our formalism should be very convenient to set limits and potential regions of discovery simultaneously, allowing scenarios where the presence of a signal is ambiguous to be studied without bias. Additionally, our method can be used to go beyond the conservative approach, and to set the strongest limit possible by exploiting the different distributions of signal and background events. With a consistent analytical method used by all dark matter direct detection experiments, the current constraints on the WIMP cross-section should be both stronger and clearer.

\section*{Acknowledgments}
JHD and CB are supported by the STFC.

 \end{document}